\documentclass[letterpaper]{JHEP3}
\usepackage{epsfig,verbatim}
\usepackage{amsmath}
\usepackage{amssymb}
\usepackage{graphicx}

\newcommand{\eref}[1]{(\ref{#1})}

\renewcommand\({\left(}
\renewcommand\){\right)}

\renewcommand\]{\right]}

\newcommand{\e}{{\rm e}}

\newcommand\eps{\epsilon}
\newcommand\mpl{m_{\rm p}}

\def\ba{\begin{eqnarray}}
\def\ea{\end{eqnarray}}
\def\be{\begin{equation}}
\def\ee{\end{equation}}

\def\O{\mathcal{O}}

\def\R{\mathcal{R}}

\def\nn{\nonumber}
\def\({\left(}
\def\){\right)}

\def\eref#1{(\ref{#1})}

\newcommand{\roughly}[1]{\mathrel{\raise.3ex\hbox{$#1$\kern-0.85em
\lower1ex\hbox{$\sim$}}}}

 \title{Hybrid
inflation with moduli stabilization and low scale supersymmetry
breaking}

\author{Sander Mooij and Marieke Postma \\
NIKHEF, Science Park 105, 1098 XG Amsterdam, The Netherlands. \\
{\emph smooij@nikhef.nl, mpostma@nikhef.nl}}

\date{\today}

\abstract {We study the supergravity hybrid inflation model of
Ref.~\cite{antusch} in the presence of a modulus field.  The
$\eta$-problem is solved by a shift symmetry for the inflaton, which
protects the inflaton mass even in the presence of the modulus field.
Inflation is (nearly) unaffected by moduli stabilization, provided the
scale of supersymmetry breaking in the post-inflation vacuum is small.
Therefore the model has the nice phenomenology that it combines low
scale supersymmetry breaking with high scale (grand unification scale)
inflation.}

\begin{document}


\section{Introduction}

Inflation, a short period of rapid expansion in the very early
universe, is by now regarded as a fundamental part of standard
cosmology, as it solves the horizon problem, the flatness problem and
the monopole problem in one go. Furthermore the mechanism of hybrid
inflation \cite{hybrid}, inflation that naturally comes to an end when
a field different from the inflaton field becomes tachyonic, has been
shown to have a natural embedding in the framework of
supersymmetry. All couplings can be chosen to be of order one (no
naturalness problem) and the field direction corresponding to the
inflaton can easily be flat enough for slow-roll inflation (no
$\eta$-problem). When moving on to local supersymmetry (supergravity)
\cite{riotto}, the $\eta$-problem may be circumvented by invoking a
shift symmetry in the K\"ahler potential, as was first proposed in
\cite{shift1,shift2}.

It is well known that inflation is a UV sensitive theory; indeed, this
is the root of the $\eta$-problem \cite{copeland,dine}. In the
language of effective field theory this can be readily seen: higher
order operators that are suppressed by some large cutoff scale can
nevertheless give a large and dominant contribution to the inflaton
mass, and thus to the $\eta$-parameter. To fully study inflation it is
therefore imperative to consider a UV completion of the theory. In
this paper we consider embedding inflation in a higher dimensional
Planck-scale theory, for example string theory. After dimensional
reduction, the 4D effective field theory will still carry traces of
its higher dimensional origin in the form of moduli fields, light
scalar fields which parametrize the shapes and sizes of the
compactified extra dimensional manifold.  For definiteness, we will
follow the seminal work of KKLT \cite{kklt} and assume that all moduli
can be fixed at some high scale by fluxes, except for the volume
modulus which is to be stabilized at a lower scale by non-perturbative
effects. The dynamics of the the volume modulus thus enters the low
energy effective field theory, and inflation should be studied in
conjunction with modulus stabilization.

There are two ways to deal with the moduli fields in the context of
inflation. The first is to make the moduli part of the inflaton
dynamics. This is for example done in racetrack \cite{race1,race2} and
K\"ahler \cite{kahlerinf,fibre} inflation models, where a modulus
field is identified with the inflaton field itself. Another approach,
the one we follow in this paper, is to decouple the physics of moduli
stabilization from the inflationary physics as much as possible. Our
set-up is as follows: we have a hybrid inflation sector and a (volume)
modulus stabilization sector, which are coupled only gravitationally
as dictated by the SUGRA action.  Even though gravitational
interactions are usually thought of as being weak, they are
generically strong enough to ruin inflation --- inflation is UV
sensitive. It has indeed been shown that ``standard'' SUSY hybrid
inflation \cite{hybridF} cannot be combined with a KKLT-like modulus
sector \cite{hybridmodF,hybridmodD} (but see \cite{hybridmodmult} for
a possible resolution). Instead we will consider a modified model of
hybrid inflation \cite{antusch}.

We want to extend the hybrid inflation model of Ref.~\cite{antusch}
with a modulus sector.  In our set-up the $\eta$-problem is solved by
a shift symmetry for the inflaton, and in addition the property that
the inflationary superpotential and its first derivative w.r.t. the
inflaton field vanishes during inflation.  Note in this respect that a
shift symmetry alone is not enough, as it is broken explicitly by the
superpotential.
This is what goes wrong in standard hybrid inflation. However, the
$\eta$-problem is not the only potential difficulty. Making sure that
the modulus field remains stabilized during inflation, implies that
the scales appearing in the SUSY breaking modulus sector are large,
resulting in a large gravitino mass. Even though the inflaton
direction is protected, large soft corrections may destabilize the
inflationary trajectory.

In this paper we describe an explicit way to stabilize the modulus
sector without running into the aforementioned troubles. The trick is
to constrain the modulus sector in such a way that its gravitino mass
is much smaller than the other scales in the problem.  Although this
presents some amount of tuning, the result is a phenomenologically
favored scenario with low scale SUSY breaking and high scale
inflation.  An explicit modulus sector that does the job is the model
developed by Kallosh \& Linde \cite{kl1,kl2} (we will refer to this as
the KL model).  

Many SUGRA or string-derived models of inflation predict a large
gravitino mass. In models based on a generic KKLT potential the
gravitino mass has to be larger than the Hubble constant during
inflation $m_{3/2} \gtrsim H_*$ \cite{kl1,kl2}, whereas in models with
a large volume compactification the bound is even stronger
$m_{3/2}^{3/2} \gtrsim H_*$ \cite{conlon}.  It has proven hard to
avoid this bound.  The KL moduli potential decouples the SUSY breaking
scale from the modulus mass, at the cost of tuning, thereby
invalidating the bound. Although it is not automatic that a KL-based
inflation scenario with low scale SUSY breaking can be constructed
\cite{hybridmodF,hybridmodD,postma}, successful models have been found
\cite{shiu}, and the model discussed in this paper is another example.
Other approaches to obtain a light gravitino can be found in
Refs.~\cite{hybridmodmult,conlon,badziak1,badziak2,badziak3}.

This paper is organized as follows. In section 2 we first briefly
describe the shift-symmetric super gravitational model of hybrid
inflation introduced in Ref. \cite{antusch}. Then, in section 3, we
explain why combining it with a generic KKLT-type modulus sector does
not work: it is impossible to find a suitable inflationary trajectory
stable in field space. In the fourth section we show that a
constrained modulus sector, of which KL is an explicit example, saves
inflation. We discuss the inflationary observables, and show some
numerical results. The paper concludes with a discussion of our
results.


\section{The model: SUGRA hybrid inflation}
We briefly describe the supergravitational shift-symmetric model of
hybrid inflation, that we want to extend by including a moduli sector
in the next sections. For a more detailed introduction we refer to the
original paper: Ref. \cite{antusch}.

The model is defined by its superpotential $W_{\rm inf}$
\be
W_{\rm inf} = \kappa\,S\left(H^2-M^2\right)+\frac{\lambda}{\Lambda}N^2H^2\,,
\label{Winf}
\ee
and K\"ahler potential $K_{\rm inf}$
\ba
K_{\rm inf} &=& |H|^2+|S|^2+\frac{1}{2}\left(N+N^*\right)^2 
+\frac{\kappa_{H}}{\Lambda^2}\,|H|^4
+\frac{\kappa_{S}}{\Lambda^2}\,|S|^4
+\frac{\kappa_{N}}{4\,\Lambda^2}\,\left(N+N^*\right)^4
\nn \\ &+&
\frac{\kappa_{SH}}{\Lambda^2}\,|S|^2|H|^2
+\frac{\kappa_{SN}}{2\,\Lambda^2}\,|S|^2\left(N+N^*\right)^2
+\frac{\kappa_{HN}}{2\,\Lambda^2}\,|H|^2\left(N+N^*\right)^2+\ldots\,.
\label{Kinf}
\ea
where the ellipses denote higher order terms, and $\Lambda$ is some
cutoff scale.  The superfield $H$ plays the role of waterfall field
responsible for ending inflation. The superfield $S$ is the so-called
driving field, as its $F$-term provides the energy density that drives
inflation. Finally, the imaginary part of $N$ is the slowly rolling
inflaton field. It is hoped that $N$ can be identified with the
right-handed sneutrino superfield, and $H$ with the grand unified
Higgs field that breaks $B$-$L$, thus providing an embedding of the
model in a grand unified theory \cite{sneutrino}. The K\"ahler
potential is invariant under a shift of $N\rightarrow N+i\mu$; this is
the before mentioned shift symmetry pivotal for keeping the inflaton
direction flat.  The superfields can be decomposed in real and imaginary
components: $H={(h_r+i h_i)}/{\sqrt{2}}$, $S={(s_r+i s_i)}/{\sqrt{2}}$
and $N={(n_r+i n_i)}/{\sqrt{2}}$.

\paragraph{Inflation}
In this model inflation takes place as the field $n_i$, the imaginary
part of $N$, slowly rolls down to a critical value $n_i^c$, while 
the other fields are in their minimum $(h_r,h_i,s_r,s_i,n_r) =
(0,0,0,0,0)$. Let us first check the stability of this minimum and then
briefly explain how inflation comes about.

During inflation the $F$-term scalar potential~\footnote{We set the
reduced Plank mass $\mpl = (8\pi G_N)^{-1/2} = 1$.}
\be
V_{F}=\e^K \[D_i W  K^{i\bar j } D_{\bar j} \bar W - 3|W|^2\],
\label{VF}
\ee 
with $D_i W = W_{i}+K_{i}\,W$ is constant:

\be V_{\rm tree} = \kappa^2 M^4, \ee 
and drives inflation with $H^2_{\rm inf} = V_{\rm tree}/3$.  All
non-inflationary fields are at an extremum of the potential
independent of the value of $n_i$. Hence the inflationary valley is a
classically and quantum mechanically stable trajectory provided all
masses squared are positive, and the mass exceeds the Hubble scale
during inflation.  The masses during inflation are
\ba
\{m_{h_r}^2,m_{h_i}^2\} &=&
\left\{ M^2 \kappa^2\(M^2\(1-\kappa_{SH}\)-2 \) + \lambda^2 n_i^4,\;
M^2 \kappa^2\(M^2\(1-\kappa_{SH}\)+2 \) + \lambda^2 n_i^4 
\right\},
\nn \\
\{m_{s_r}^2,m _{s_i}^2 \} &=&
\{-4M^4 \kappa^2\kappa_S,\; -4M^4 \kappa^2\kappa_S\},\nn \\
\{m_{n_r}^2,m_{n_i}^2\} &=&
\{2M^4 \kappa^2\bigl(1-\kappa_{SN}\bigr),\; 0 \}.
\label{mass}
\ea
Here, and from now on, we set $\Lambda=1$.  We see that $h_r$ becomes
tachyonic when $n_i$ drops below the critical value $(n^{2}_i)^c
\approx \sqrt{2} \kappa M/\lambda$. This will mark the end of
inflation.  The $s_r$, $s_i$ and $n_r$ directions are stable as long
as $\kappa_{SN}<\frac{5}{6}$ and $\kappa_S < -\frac{1}{12}$. This is
one of the reasons for including the higher order terms in the
K\"ahler potential \eref{Kinf}. (The other is that the inflationary
observables depend on $\kappa_{SH}$ \cite{antusch}; taking
$\kappa_{SH} = \O(10)$ the spectral index can be brought closer to the
WMAP central value.)

Thanks to the shift symmetry, $n_i$ itself does not acquire any mass,
independent of the higher order terms in the K\"ahler potential; at
tree level it is a flat direction in field space. The slow roll
parameter $\eta = V''/V $ is small, with prime denoting derivative
w.r.t. the canonically normalized inflation field, and there is no
$\eta$-problem.  This is in contrast with ``standard'' SUSY hybrid
inflation \cite{hybridF} where higher order terms lift the flatness of
the potential, and thus must be tuned \cite{riotto}.  The reason for
this marked difference is that in our model $W_{\rm inf}$ vanishes
during inflation (as well as many first and second derivatives of
$W_{\rm inf}$), thereby killing all possible inflaton mass terms.  In
standard SUSY hybrid inflation on the other hand $W_{\rm inf} \neq 0$,
and the $\eta$-problem resurfaces despite the shift symmetry.  It is
this remarkable property of the inflaton superpotential that led the
authors of \cite{antusch} to suggest that the model can be combined
with a modulus sector.  In the next sections we will take a closer
look at this claim.

The inflaton potential is generated by the 1-loop Coleman-Weinberg
potential \cite{CW}, from the mass splitting between fermions and
bosons.  Only the waterfall fields have inflaton-dependent mass terms
and contribute to the inflaton potential. Writing the mass of the
waterfall fields and their fermionic superpartners in the form
$m^2_{h_{r,i}} = \mu^2(x^2+y^2\pm 1)$ and $\tilde m^2_{h_{r,i}} = \mu^2 x^2$
with
\be
\mu^2 ={2\kappa^{2}M^{2}}
,\quad
x = \frac{\lambda^2 n_{i}^{4}}{2\kappa^{2}M^{2}},
\quad
y =\frac{M^{2}}{2}(1-\kappa_{SH}),
\label{xy1}
\ee
then the loop potential is given by \eref{Vloop} in appendix
\ref{A:CW}. We now have the effective potential
\be V_{\rm inf}=V_{\rm tree} + V_{\rm loop}(n_i).  
\label{Vtot}
\ee
The $n_i$-direction in field space, flat at tree-level, gets slightly
lifted at the one-loop level. With a suitable choice of parameters
this effective potential can generate inflation. The inflaton field
$n_i$ slowly rolls down until it reaches the critical value $n_i^c$
where inflation ends.  We note that for $y^2 <0$, or equivalently
$\kappa_{SH} > 1$, the CW-potential has a maximum at $n_{i}^{\rm max}$
given in \eref{xmax}.  This introduces a constraint on the initial
field value of the inflaton field which has to be smaller than
$n_i^{\rm max}$, to make sure that the inflaton rolls towards the
``right'' minimum.  On the other hand for $y^2 < 0$, the loop
potential steadily increases with $n_i$, and there is no such problem.

With the potential \eref{Vtot} one can calculate the inflaton value
$(n_i)_*$ at horizon-exit, 60 e-folds before the end of inflation, when
observable scales leave the horizon. Here the slow-roll parameters
$\epsilon$, $\eta$ and $\xi^2 $ can be evaluated, and consequently the
power spectrum $\mathcal{P}_{\mathcal{R}}$, the scalar spectral index
$n_S$, the tensor-to-scalar ratio $r$, and the running of the scalar
spectral index ${d n_S}/{d \log{k}}$.

\paragraph{After inflation}
When the inflation field $n_i$ reaches its critical value $n_i^c$, the
waterfall field $h_r$ becomes tachyonic. Inflation ends with a phase
transition during which the waterfall field obtains a non-zero vev.
The post-inflationary vacuum field values are
$\{h_r,h_i,s_r,s_i,n_r,n_i \}= \{\pm \sqrt{2} M,0,0,0,0,0\}$, and $V
=0$ corresponding to zero cosmological constant.

\paragraph{Numerical results}
In Ref. \cite{antusch} it is shown that in the parameter space 
\be
\biggl( 
\kappa=\mathcal{O}(10^{-1}),M=\mathcal{O}(10^{-3}),
\lambda=\mathcal{O}(10^{-1}),
\kappa_{SH}=\mathcal{O}(1\!-\!10)\biggr)
\ee
many solutions can be found that satisfy the WMAP $1\sigma$ range for
the power spectrum
$\mathcal{P}_{\mathcal{R}}^{\frac{1}{2}}=(5.0\pm0.1) \times 10^{-5} $
and scalar spectral index $n_s=0.960^{+0.014}_{-0.013}$
\cite{WMAP5}. The tensor to scalar ratio $r$ typically becomes of
order $(10^{-5})$ which easily satisfies the WMAP bound $r<0.2$. The
model fails on the prediction of ${d n_s}/{d \log{k}}$: it
typically predicts a value of order $(10^{-4})$ while WMAP measured
$-0.0032^{+0.021}_{-0.020}$. As the accuracy of this measurement is
rather low, this does not seem to be a serious problem.


\section{Adding the modulus sector}

The inflaton model described in the previous section is an effective
theory, arising as a low-energy effective description of an underlying
Planck scale theory.  If the UV completion is an extra dimensional
theory, we expect moduli fields to appear in the 4D effective action.
The moduli fields parametrize the sizes and shapes of the extra
dimensions.  In case the vacuum manifold is degenerate, the moduli
correspond to massless modes appearing in the low energy effective
four-dimensional theory.

For definiteness we concentrate in this paper on a KKLT type moduli
sector \cite{kklt}, arising from compactifications in type IIB string
theory.  KKLT showed that all complex structure moduli (shape moduli)
can be stabilized by fluxes. In the simplest case there is only one
K\"ahler modulus (size modulus) left, the volume modulus, which
appears in the 4D effective theory.  This modulus, in turn, is
stabilized by invoking non-perturbative effects, coming from either
gaugino condensation or instantons. Finally, to arrive at a zero
cosmological constant, a non-supersymmetric uplifting term is added,
generated by an anti-D3 brane located at the bottom of a throat in the
compactification manifold.

To combine the inflaton and modulus sector we simply add their
respective K\"ahler- and superpotentials~\footnote{The K\"ahler
potential does not have to be separable in modulus and inflaton
field, e.g. $K_{\rm inf}$ can appear inside the log. We checked that
its exact form does not affect our qualitative results.}:
\be
W = W_{\rm inf} + W_{\rm mod}, \qquad
K = K_{\rm inf} + K_{\rm mod}
\ee
with
\be
K_{\rm mod} = - 3 \ln(T + \bar T) .
\label{modelT}
\ee
For the moment we ignore corrections to this tree level K\"ahler potential. These will be treated in section 4.1.
\subsection{General approach}
The function $W_{\rm mod}(T)$ generically contains a constant term
$W_0$ arising from integrating out the stabilized moduli and a
non-perturbative potential 
that is to stabilize
$T$. In this section we will work with a generic function $W_{\rm
mod}(T)$ and see what restrictions on this function we get to make
inflation work in this moduli-extended framework. We choose a KKLT
uplifting potential $V_{\rm up} = {c}/{(T+\bar{T})^2}$, with $c$ a
constant tuned to solve the cosmological constant problem. However,
its specific form is not so important for our discussion.

As before the modulus fields can be decomposed in real and imaginary
parts: $ T= \sigma + i \alpha$. Choosing the phases in the
superpotential judiciously, we can set $\alpha = 0$ to zero
consistently. We define $\sigma = \sigma_0$ at the minimum of the
F-term modulus potential in the absence of the inflaton sector, i.e.
\be 
\partial_\sigma V^F_{\rm mod}\big|_{\sigma = \sigma_0} = 0
\quad \Leftrightarrow \quad
D_T W \big|_{\sigma = \sigma_0} = 0.
\label{sigma0}
\ee
Due to the uplift term and the presence of the inflaton sector,
$\sigma$ is displaced from its F-term minimum both during and after
inflation. If the displacement is minimal the inflationary trajectory
and the post-inflation minimum are only slightly affected as well, and
the moduli sector may be combined with inflation. In this case $\sigma
\approx \sigma_0$ and $D_T W \approx 0$ are still good approximations.
In the rest of this section we discuss the general conditions the
moduli sector has to satisfy for this to be the case, followed --- in
the next section --- by an explicit example.  There are many
pitfalls. When a modulus sector is included, the $\eta$-problem may
reappear, the vacuum after inflation and/or the inflationary
trajectory may be destabilized, and the corrections to the waterfall
fields may hamper a successful exit to inflation.

\paragraph{$\eta$-problem}

We have seen that in the absence of a moduli sector the tree level
inflaton mass is zero, as a consequence of the shift symmetry and the
fact that $W_{\rm inf} = 0$.  Due to the shift symmetry the K\"ahler
potential is independent of $n_i$, and thus any mass for $n_i$ must
come from the second derivative of the term in square brackets in
\eref{VF}. The fact that the modulus superpotential is non-zero does
not change the results, the inflaton potential is still flat at tree
level. All terms in $m_{n_i}^2$ proportional to $W_{\rm mod}$ or its
derivatives are multiplied by $(W_{\rm inf})_{n_i n_i}$, which is zero
during inflation.  As the $\eta$-problem is usually the main obstacle
to embedding inflation in a supergravity theory, this is no small
feat.

\paragraph{Stability of the vacuum}

Consider the vacuum after inflation. We suppose the post-inflationary
minimum to occur at $\{h_r,h_i,s_r,s_i,n_r,n_i,\sigma ,\alpha\}= \{\pm
\sqrt{2} M,0,0,0,0,0,\sigma_0,0\}$.  For the post-inflation scalar
potential we find
\be V_{\rm vac} = \frac{c}{(4\sigma)^2}+ V^F_{\rm mod} + f(M^2),
\qquad 
V^F_{\rm mod}= \frac{-3W_{\rm mod}^2
+ \frac43 \sigma^2 (D_T W_{\rm mod})^2}{(2\sigma)^3} ,
\label{Vpi}
\ee
where each term in $f(M^2)$ is either proportional to $D_T W$ or to
$W$ (with $D_T W =-{3 W}/{(2 \sigma)} + W_T$). For parameters
that keep the modulus stabilized during inflation (discussed below),
the $M$-dependent corrections $f(M^2)$ to the modulus potential after
inflation are small, and do not destabilize the potential minimum.

The first derivatives of the scalar potential with respect to the
eight real fields, evaluated at the postulated potential minimum after
inflation, are manifestly zero or involve again small functions of
$M^2$ proportional to $D_T W$ or $W$ indicating that the minimum of
some of the fields is slightly displaced. One of the displaced fields
is the modulus field, which is shifted from its F-term potential
minimum $\sigma_0$ due to the presence of the uplift term. This shift
is typically small.

Second derivatives again involve many functions of $D_T W$, $W$ and
$c$. The vacuum mass of the field $n_r$ is most seriously affected by
moduli corrections, and runs the risk of going tachyonic:
\be
m_{n_R}^2 \big|_{\rm vac} =
\frac{4 (D_T W_{\rm mod})^2 \sigma^2 - 3 W_{\rm mod}^2+ \O(M^2)}{12 \sigma^3} .
\label{nr}
\ee
Indeed, for $D_T W_{\rm mod} \approx 0$ the mass is tachyonic unless
$W_{\rm mod} \lesssim M$ is sufficiently small, and the $\O(M^2)$
terms dominate.

\paragraph{Stability during inflation}

The tentative inflationary trajectory is
\be
\{h_r,h_i,s_r,s_i,n_r,n_i,\sigma ,\alpha \}
=\{0,0,0,0,0,n_i,\sigma_0,0\}.
\label{trajectory}
\ee
 We have to check whether
this is still an extremum when the modulus potential is turned on.  As
before $n_i$ is the slowly rolling inflaton field.  The potential
during inflation along this trajectory is then
\be V_{\rm inf} = 
\frac{c}{(4\sigma)^2}+ V^F_{\rm mod}+ \frac{\kappa^2 M^4}{(2\sigma)^3},
\label{Vi}
\ee 
with $V^F_{\rm mod}$ defined in \eref{Vpi}. If $\sigma \approx
\sigma_0$ the first two terms in the above expression nearly cancel,
and the last term is as before the energy density driving inflation.
However, this energy density is now modulus dependent. If this term is
too large, the displacement in $\sigma$ is large, or worse, the barrier
separating in the potential disappears and $\sigma$ rolls off to
infinity. 

The fields are all at an extremum for the inflationary trajectory,
except for the modulus and the $s_r$ field. The non-vanishing first
derivatives are
\ba
\partial_{s_r} V_{\rm inf}&=&
\frac{\kappa M^2 ((D_T W_{\rm mod})\sigma +W_{\rm mod})}{2 \sqrt{2} \sigma^3},
\label{dvp}  \\
\partial_{\sigma} V_{\rm inf}&=&
\frac{-3(3\kappa^2 M^4+4 c\sigma) 
+ 8\sigma^2 (D_T W_{\rm mod}) (-2 D_T W_{\rm mod}
-\frac{3 W_{\rm mod}}{\sigma}+\sigma W''_{\rm mod})}{24 \sigma^4}, \nn
\ea
where primes denote derivatives with respect to $\sigma$. We see
indeed that during inflation the minimum of the $\sigma$-field does
not occur at exactly $\sigma=\sigma_0$: now it is both the uplift and
the inflationary energy density that shifts the minimum away. In
addition, the field $s_r$ is not minimized at $s_r =0$.  For $D_T
W_{\rm mod} \approx 0$, the first derivative is proportional to $W_{\rm
mod}$ and is typically large. This can have dramatic consequences, as
we will see. The matrix of second derivatives evaluated at the
inflationary minimum is not diagonal anymore, as $V_{s_r \sigma}$ does
not vanish. This coupling between $s_r$ and $\sigma$ could already be
foreseen from \eref{dvp}. We also find a similar coupling between
$s_i$ and $\alpha$, but as they both have their minimum at zero this
coupling will not have any significant consequences.

Just as in the case without moduli fields ({\it cf}. the discussion
below \eref{mass}), we need some tuning of the $\kappa$-parameters in
the inflationary K\"ahler potential to maintain positive definite
masses squared.  Since expressions are long, we only explicitly give
the mass of the field $h_r$, that can be compared to \eref{mass}
\ba
V^{\rm inf}_{h_r h_r}=
\frac{1}{(2\sigma_{0})^{3}} &\bigg [&
\kappa^2 M^2(\bigl(M^2(1-\kappa_{SH})-2\bigr) + \lambda^2 n_i^4 
-2 W_{\rm mod}^2 + 2\sigma \lambda n_i^2 D_T W_{\rm mod} 
\nn \\
&&+ W_{\rm mod}\bigl(\lambda n_i^2 - 4\sigma D_TW_{\rm mod}\bigr) 
+\frac{4}{3}\sigma D_T W_{\rm mod}\bigl(\sigma D_T W_{\rm mod} + 3\bigr)
\bigg].
\label{mhr2}
\ea
Once again, for $D_T W_{\rm mod} \approx 0$, the corrections --- in
this case to the waterfall masses --- scale with $W_{\rm mod}$ and
are potentially large.

\paragraph{Waterfall mechanism and CW-loop}
The above expression becomes much more complicated when we take the
displacement of $s_r$ into account, see \eref{dvp}. If we allow $s_r$
to be non-zero we find among many other terms
\be
\delta m_{h_{r,i}}^2 = \frac{\kappa^2 s_r^2}{2\sigma^3} + ... \label {hrs}
\ee 
This indicates that the shift in $s_r$ can do a lot of harm to our
model. Once the waterfall masses get dominated by terms like
\eref{hrs}, the Coleman-Weinberg loop potential changes drastically and
inflation is no longer possible. Therefore, we absolutely need the
displacement in $s_r$ to be small.

\subsection{Discussion}

As discussed, for a generic superpotential $(D_T W_{\rm mod}) \approx
0$ as this minimizes the F-term superpotential \eref{sigma0}, and
corrections to the inflaton potential scale with $W_{\rm mod}$ (which
is the only scale in the moduli sector).  $W_{\rm mod}$ should be
large enough to assure the modulus remains stabilized during
inflation, yet small enough to ensure that the vacuum and inflationary
trajectory is not destabilized. This does not seem to be easy. And
indeed, for a KKLT modulus sector, which is of the above described
generic form, this is impossible.  In the original KKLT paper
\cite{kklt} the non-perturbative potential is a single exponent, and
the superpotential is
\be W_{\rm mod}=-W_0 + A e^{-a T}.  \ee 
where the sign in front of $W_0$ is chosen such that the potential is
minimized by $\alpha =0$.  The minimum of $V_{\rm vac}$ occurs for $D_T
W_{\rm mod} \approx 0$ and $W_{\rm mod} \sim W_0$.  Let us go through
all modulus corrections for this specific choice of superpotential.

For the $n_r$-direction to be stable in the vacuum after inflation,
see \eref{nr}, we have to demand $W_0 \lesssim \kappa M^2$.  From the
perspective of the modulus field, the inflationary energy density
$\kappa^2 M^4$ acts as an additional uplift term \eref{Vi}.  If this
term is too large, $\sigma$ is destabilized.  To avoid this one needs
$\kappa^2 M^4/ (2\sigma)^3 \lesssim V_{\rm up}$ or 
\be
\kappa M^2 \lesssim W_0
\label{constraint1}
\ee
It follows that stabilizing the modulus during inflation plus
stabilizing the vacuum are both possible only for a very limited range
of parameters: $W_0 \approx \kappa M^2$.  But what kills the KKLT
model are the corrections it gives to the waterfall fields.  As
anticipated from \eref{dvp} it follows that both $s_r$ and $\sigma$
are displaced considerably during inflation. Numerically we find $s_r
\sim \O(10^{-1} - 10^{-2})$ (where we used $V_{\rm inf}^{1/4} \sim
M_{\rm GUT}$).  The corresponding correction to the waterfall field
\eref{hrs} is enormous, hampering a graceful exit to inflation.

How to salvage inflation?  Taking a look at the modulus corrections
(\ref{nr}, \ref{Vi}, \ref{dvp}, \ref{hrs}), we see they all vanish in
the limit that both 
\be
D_T W_{\rm mod} \big|_{\sigma =\sigma_0} \approx 0
\qquad \& \qquad
W_{\rm mod}  \big|_{\sigma =\sigma_0} \approx 0.
\label{constraint2}
\ee
The first condition is assured by minimizing of the F-term potential
\eref{sigma0}, but the second constitutes an extra constraint on the
modulus potential which can be satisfied by tuning the parameters in
the superpotential. Such a tuning is not possible for the one-exponent
KKLT model. Kallosh \& Linde (KL) constructed a modulus sector with
two exponents, with the parameters carefully tuned, such that
\eref{constraint2} is satisfied \cite{kl1, kl2}.  We will discuss this
model in detail in the next section.  The only constraint left is then
\eref{constraint1}, assuring that the modulus remains fixed during
inflation.

The fine-tuning required to set $W_{\rm mod} \approx 0$ is the same
tuning that creates a hierarchy between the gravitino and modulus mass
with $m_{3/2} \ll m_T$.  Since $H_* \gtrsim m_T$ (from
\eref{constraint1}), this tuning allows to have low scale SUSY
breaking with high scale inflation --- something that seems impossible
in non-fine tuned models. This was the motivation behind the KL model.
Note that since in hybrid inflation $V_{\rm inf} \sim M_{\rm GUT}^4$,
without this tuning, it is impossible to get the phenomenologically
favored TeV scale SUSY breaking.

Finally we would like to contrast the results with standard SUSY
hybrid inflation \cite{hybridF,hybridmodF,hybridmodD}.  In the
standard case, the $\eta$-problem reappears once a modulus sector is
included; the reason is that in these models the inflaton
superpotential is non-zero $W_{\rm inf} \neq 0$, and many terms mixing
the modulus and inflaton sector appear in $V_F$.  In addition, the
waterfall masses get large corrections, just as we found above.
Although each of these problems can be solved separately by a
fine-tuned condition on the modulus potential, they cannot be solved
simultaneously.  Since in our case, the $\eta$-problem has dropped off
the list, inflation can be rescued by a single tuning.


\section{Inflation with a KL modulus sector}

As discussed in the previous section, hybrid inflation may be combined
with a modulus sector provided the latter satisfies
\eref{constraint2}.  In this section we work out the details,
focusing on the KL modulus sector introduced by Kallosh \& Linde in
\cite{kl1,kl2}. Augmenting the KKLT potential by an additional
non-perturbative exponential factor, it is possible (by tuning the
parameters) to construct a SUSY Minkowski minimum with $D_T W = W_T
=0$. The superpotential is
\be 
W_{\rm mod} = -W_0 + A \e^{-a T} - B \e^{-b T} 
\ee
with $W_0$ and $\sigma_0$:
\be
W_0 = w_0 \equiv A \( \frac{bB}{aA} \)^{a/(a-b)}
- B \( \frac{bB}{aA} \)^{b/(a-b)}, \qquad 
\sigma_0= \bar \sigma_0 \equiv \frac{1}{a-b} \ln\(\frac{a A}{bB}\).
\ee
So, at the cost of fixing $W_0$ and introducing another exponent in
the non-perturbative potential, we now explicitly have $D_T W = W_T
=0$, and thus $V^F_{\rm mod}=0$, in the vacuum after inflation
$\{h_r,h_i,s_r,s_i,n_r,n_i,\sigma ,\alpha\}= \{\pm \sqrt{2}
M,0,0,0,0,0,\sigma_0,0\}$. No uplift term is needed, and SUSY is
unbroken.

We can get a small but non-zero gravitino mass by perturbing the
SUSY Minkowski solution
\be
W_0 = w_0 + \eps_w.
\label{pert}
\ee
As long as the perturbation is small enough $D_T W_{\rm mod} \approx
0$, $W_{\rm mod} \approx \eps_w$ and \eref{constraint2} is still
satisfied. We will determine below how small $\eps_w$ has to be.  With
this perturbation the minimum of the F-term potential, located at
$\sigma_0 = \bar \sigma_0 + \O(\eps_w)$, is SUSY AdS, and a small
uplift $V_{\rm up} \approx 3 W_0^2/(2\sigma_0)^3$ is needed to get
zero cosmological constant. SUSY is broken in the process. In this
set-up there is a large hierarchy between the gravitino $m_{3/2} =
\e^K |W| \propto \eps_w$ and modulus mass $m_\sigma \propto
\sqrt{V_{\sigma \sigma}} \propto W_0$.
%
%

\subsection{Inflation}

Let us see how inflation works for the hybrid inflation model
described in section 2 combined with a KL modulus sector.

\paragraph{Stability of the vacuum}

Since the function $f(M^2) \sim \eps^{2}$ in \eref{Vpi} the inflaton
corrections to the modulus minimum after inflation are small.
Likewise the modulus correction to the inflaton sector are small.  The
mass of $n_{r}$ in \eref{nr} is manifestly positive definite in the
vacuum. We checked numerically the stability of the vacuum.

\paragraph{Stability during inflation}

From \eref{Vi}  we see that during inflation we now have
\be V_{\rm inf}  \approx
\frac{\kappa^2 M^4 + \O(\eps_{w}^{2})}{(2\sigma_{0})^3}.
\ee 
The inflationary trajectory is slightly shifted from the tentative
inflationary trajectory \eref{trajectory}, as the first derivatives
$V_{i}$ with $i =\{s_{r},\sigma\}$ are non-zero \eref{dvp}. Expanding
in small $\eps_{w}$ this shift is
\ba
\delta \sigma &=&-\frac{9 \kappa^2 M^4 
\bigl(3-4\kappa_S\bigr)}{4 a^2 b^2 \kappa_S  W_0^2}+
\eps_{w} \frac{3(1-2\kappa_S)}{4 a b \kappa_s W_0 \sigma_0},
\nn \\
\delta s_r &=& -\frac{9 \kappa M^2}{8 \sqrt{2} a b W_{0} \kappa_{S} \sigma_{0}^2}
- \eps_{w}\frac{1}{\sqrt{2} \kappa \kappa_{S} M^2}.
\label{deltazooi}
\ea
The shift due to the inflation sector, which is the $\eps_{w}$
independent part, is small, and harmless for inflation.  The
corrections due to the modulus sector scale with $\eps_{w}$ and can be
larger depending on the size of $\eps_{w}$.  The mass matrix is nearly
diagonal. Except for the $s_r$-field, the masses for all the inflaton
fields are as before \eref{mass}, up to an overall scaling by
$(2\sigma_0)^3$, and up to order $\delta m_i^{2} =
\O(\eps_{w}^2/(2\sigma_{0})^{3})$ corrections. From the masses of
$s_r$ and $n_r$ one can deduce constraints on $\kappa_{SN}$ and
$\kappa_S$, just as we did before around \ref{mass}:
\be m^2_{n_r}=\frac{2 \kappa^2 M^4
(1-\kappa_{SN})}{(2\sigma_0)^3}, \qquad m^2_{s_r}=\frac{\kappa^2 M^4
(3-4\kappa_S)}{(2\sigma_0)^3}
\ee 
lead to the constraints $\kappa_{SN} < \frac{5}{6}$ and $\kappa_S <
\frac{2}{3}$.  It follows that for $\eps_{w} \gtrsim \kappa M^{2}$ the
moduli corrections dominate, and one of the masses, depending on the
choice of $\kappa_{i}$ parameters in the K\"ahler potential
\eref{Kinf}, may go tachyonic, thereby destroying inflation.

\paragraph{Waterfall mechanism and CW-loop}

A stronger bound on the value of $\eps_{w}$ may be obtained by looking
at the waterfall masses.  Writing the masses of the bosonic waterfall
fields and their superpartners in the form $m^2 = \mu^2(x^2 +y^2 \pm
1)$ and $\tilde m^2 = \mu^2 x^2$, we find
\be
\mu^2 = \frac{2\kappa^2 M^2}{(2\sigma_0)^3} + \O(\eps_w), \quad
x^2 = \frac{\lambda^2 n_i^4}{2\kappa^2 M^2} + \O(\eps_w),\quad
y^2 = \frac{M^2}{2}(1-\kappa_{SH}) - 
\eps_w^2 \frac{\kappa_{SH} \lambda^2 n_i^4}{4 \kappa^2 \kappa_S^2 M^2}
\label{xy2}
\ee
where the dominant moduli correction for our purposes is the
$\O(\eps_w^2)$ correction in $y^2$.  Although $y^{2} \ll x^2$, and is
an unimportant contribution to the waterfall field mass, it is
relevant for the 1-loop potential. As explained in the appendix
\ref{A:CW}, the reason is that the dominant terms cancel between the
bosons and the fermions in the Coleman-Weinberg potential \eref{CW}.
Indeed, even in the absence of moduli corrections the term $\propto
M^{2}(1-\kappa_{SH})$ in the boson mass causes the loop potential to
develop a maximum for $\kappa_{SH} >1$ \eref{xmax}. If the
$\eps_w^2$-correction in $y^2$ dominates, the loop potential steepens
for large $n_i$. In the case of $\kappa_{SH} >0$, this results in the
maximum shifting to smaller values of $n_{i}$, until at some point it
becomes impossible to get 60 e-folds of inflation. In the opposite
limit $\kappa_{SH} <0$ it results in a larger spectral index, in
contradiction with observations.  Either way, inflation is ruined if
the moduli corrections get too large.  Using \eref{xy2} this gives the
bound
\be
\eps_{w} \lesssim  0.1 -0.01 \kappa M^{2} 
\label{epsbound}
\ee
where the exact value depends on the $\kappa_{i}$ values, and the
precise parameters.  This estimate is confirmed by our numerical
calculation.

\paragraph{Corrections to modulus K\"ahler potential}

What remains to address is an analysis of possible corrections to the tree-level K\"ahler moduli potential introduced in \eref{modelT}. Denoting $\alpha'$ \cite{Becker} and $g_s$ \cite{Berg,Cicoli} corrections by $\theta_1$ and $\theta_2$ respectively, we replace \eref{modelT} by
\be
K_{\rm mod}=-2\ln\left[\left(T+\bar{T}\right)^{3/2} + \theta_1\right] + \frac{\theta_2}{\left(T+\bar{T}\right)^{3/2}}.
\ee
In the natural limit $\theta_1<\left(T+\bar T \right)^{3/2}$ we can expand the logarithm to arrive at
\be
K_{\rm mod}=-3\ln\left(T+\bar{T}\right) +  \frac{-2\theta_1+\theta_2}{\left(T+\bar{T}\right)^{3/2}} \equiv -3\ln\left(T+\bar{T}\right) +  \frac{\theta}{\left(T+\bar{T}\right)^{3/2}},
\ee
where we have defined $\theta\equiv -2\theta_1+ \theta_2$.\newline
\newline
To check analytically whether these corrections might spoil inflation, we calculate the induced displacement in the minimum of the $s_r$ field. Before we have seen that large $s_r$ contributions ruin the waterfall mechanism and CW loop effect, so this seems the right quantity to consider.\newline
In leading order in $M^2$ we find that these contributions are naturally small:
\be
\delta s_r= \delta s_r\rvert_{\theta=0} \left(1+\frac{\theta}{\left(T+\bar T \right)^{3/2}}\right).
\ee
This already suggests that K\"ahler corrections leave the model unaffected. Indeed, this is confirmed by our numerical analysis. As long as $\theta<\left(T+\bar T \right)^{3/2}$ all results of the previous sections apply: there is no danger for the K\"ahler corrections to spoil the model.

\subsection{Numerical analysis}

\begin{figure}
\begin{center}	
  \hspace*{-.5cm}
 \leavevmode\epsfysize=4.5cm \epsfbox{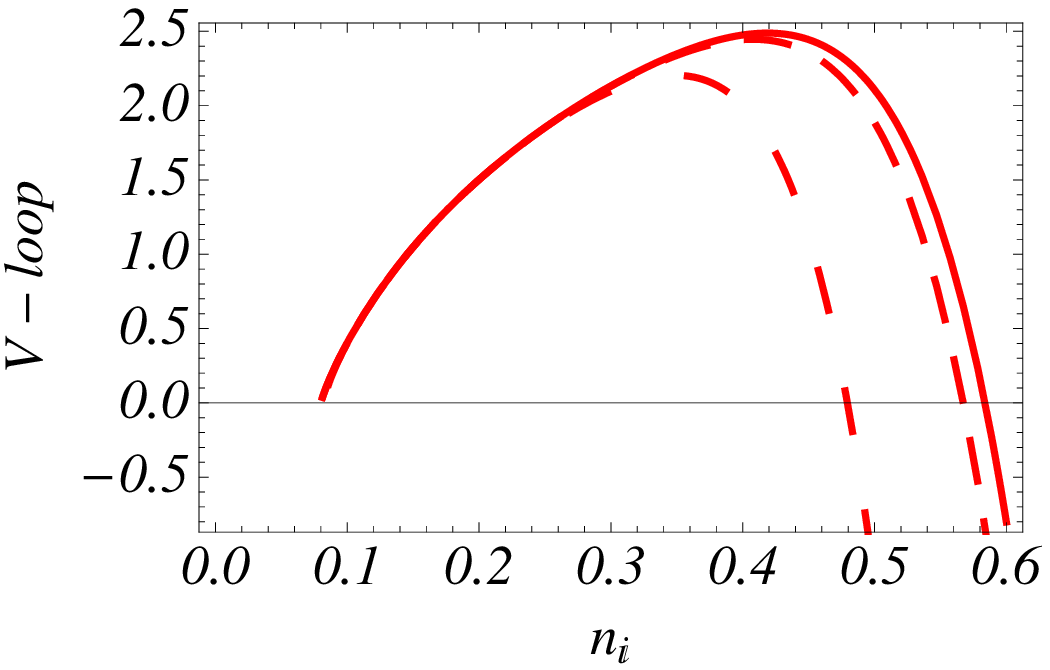}
	\hspace*{.5cm}
	\leavevmode\epsfysize=4.5cm \epsfbox{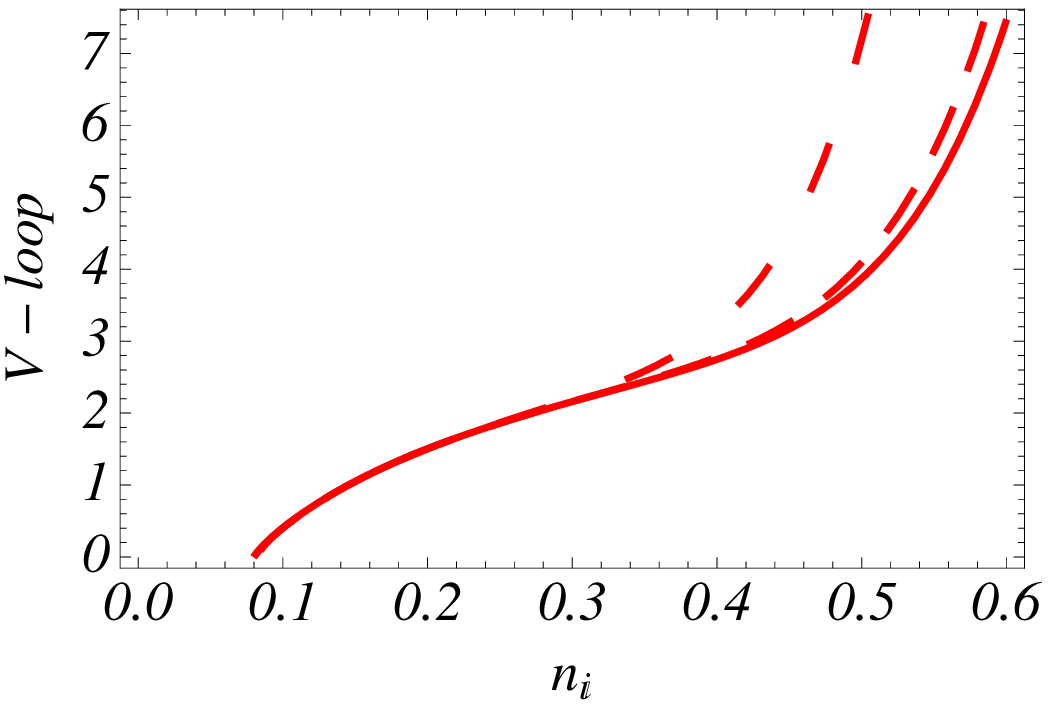}
 \end{center}
\caption{The Coleman-Weinberg potential, rescaled by a factor
$10^{15}$, as a function of $n_{i}$ for
$\eps_{w}=\{0,\,10^{-7},\,10^{-5} \}$ corresponding to the solid,
short dashed and dashed lines respectively.  On the left are the
results for $\kappa_{SH} =1$, on the right for $\kappa_{SH} =-1$.}
	\label{F:Vloop}
\end{figure}

Adding a modulus sector to inflation, the F-term potential and thus
all masses squared are rescaled by a factor $\e^K = (2\sigma)^{-3}$.
We can absorb this factor in the parameters of the superpotential via
\be 
\bar \kappa = \frac{\kappa}{(2\sigma_0)^{3/2}},\quad 
\bar \lambda = \frac{\lambda}{(2\sigma_0)^{3/2}},\quad 
\bar A =  \frac{A}{(2\sigma_0)^{3/2}},\quad 
\bar B =\frac{B}{(2\sigma_0)^{3/2}},\quad 
\bar W_0 =\frac{W_0}{(2\sigma_0)^{3/2}}.
\label{rescale}
\ee 
It is the barred quantities that give the effective couplings between
the fields, and that can be measured (in principle) in
experiments. The rescaling allows to easily compare the parameter
space for hybrid inflation without moduli as described in Section 2
and discussed in detail in Ref. \cite{antusch}, with the set-up where
a modulus potential is included.  If in the former case the model
gives the right predictions for the density perturbations for a given
set of parameters, for example $\{\kappa =0.14, M =0.003,...\}$, the
same observational results are obtained in the setup up with a modulus
field if we choose the same numerical values for the barred quantities
$\{\bar \kappa =0.14, M =0.003,...\}$. This correspondence works up
to $\O(\eps_w)$ corrections.  We checked numerically that with the
above identification we get the same parameter space for successful
inflation, e.g. including the same $\kappa_{SN}$ dependence, as found
in Ref. \cite{antusch}.

Consider an explicit numerical example. For the inflaton sector we
choose parameter values
\be
\bar \kappa=0.14, \quad M=0.003,\quad \bar \lambda=0.1, \quad \kappa_{SH}=1,
\label{kM}
\ee
and all other $\kappa_i$ equal to $-1$.  As discussed in section 2
this assures stability of the inflationary trajectory.  For the
modulus sector we take
\be
\bar A = 1,\quad \bar B =1.03, \quad a =\frac{2\pi}{100}, \quad 
a =\frac{2\pi}{99}.
\label{AB}
\ee
which gives $W_0 = 0.276 + \eps_w$ and $\sigma_0 = 62.41$.  The exact
parameter values in \eref{AB} are not so important, what matters is
the resultant value for $W_0$ and to some lesser extent $\sigma_0$.

\begin{figure}
     	\begin{center}
 	\leavevmode\epsfysize=4.5cm \epsfbox{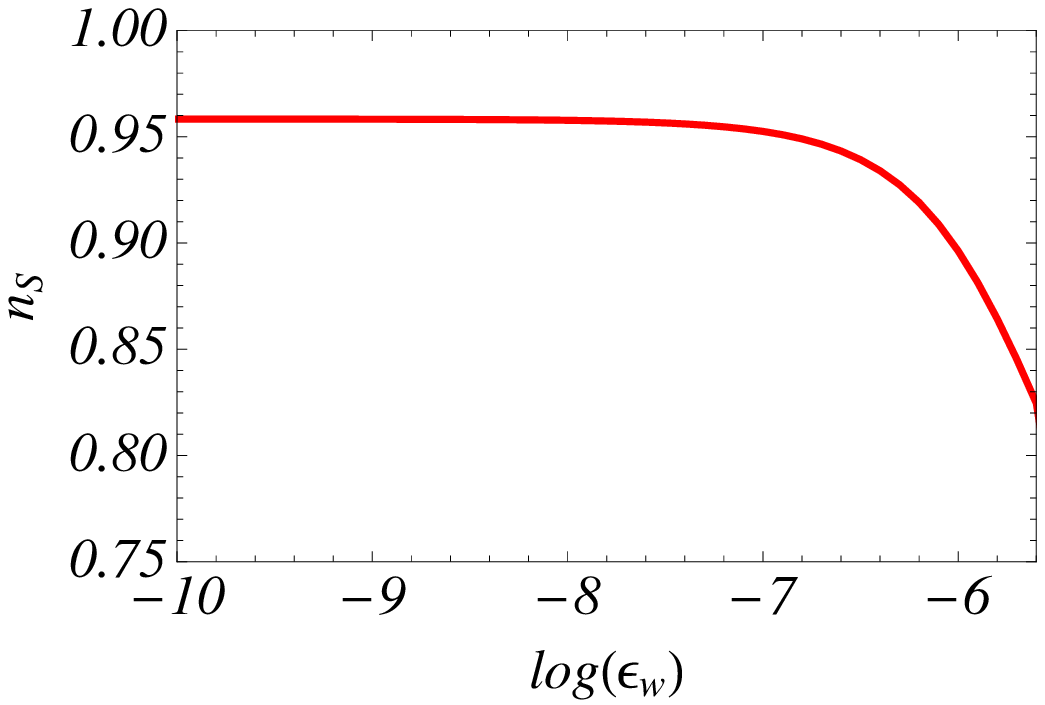}
	\hspace*{.5cm}
	\leavevmode\epsfysize=4.5cm \epsfbox{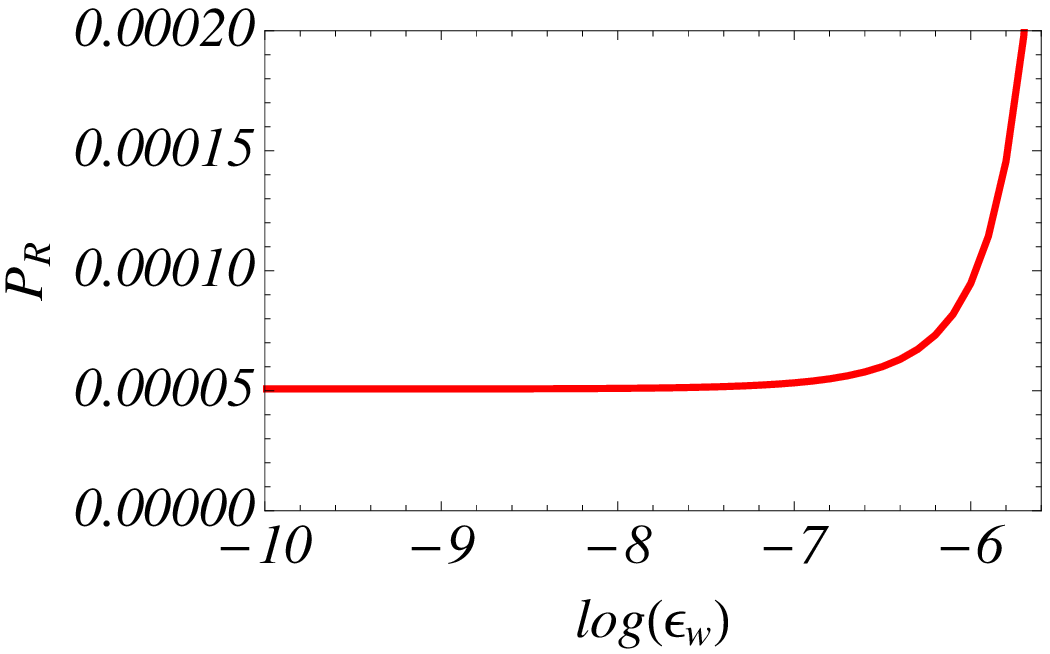}
	\end{center}
	\caption{The spectral index $n_{s}$ (left) and power spectrum $P_{\R}$ (right) as a function of $\eps_{w}$ for the parameters mentioned in the text.}
	\label{F:ns}
\end{figure}

As anticipated, as we increase $\eps_w$ we see that the moduli
corrections first appear in the loop potential.  For $\eps_w =
\{0,\,10^{-7},\,10^{-5}\}$ we get $\delta \sigma =
\{2.06,\,2.39,\,5.33\} \times 10^{-5}$ and $\delta s_r
=\{3.26,\,3.66,\,7.28\} \times 10^{-4}$.  These values match our
estimates \eref{deltazooi}.  Although the increase in $\delta s_r$
seems quite moderate, the effects are nevertheless visible in the loop
potential, where the maximum is shifting to increasingly small
$n_i$-values. This is plotted in Fig.\ref{F:Vloop}. For $\eps =
10^{-5}$ the loop potential gets modified in such a way that the
slow-roll trajectory is not large enough to accommodate 60 e-folds of
inflation.  Fig. \ref{F:Vloop} also shows the equivalent results for
$\kappa_{SH} = -1$ and for the rest the same parameters; now the
potential steepens to fast for $\eps_w > 10^{-5}$ pushing the spectral
index to values larger than one.

Fig.~\ref{F:ns} shows the spectral index and power spectrum as a
function of $\eps_w$. For small $\eps_w$ the results are identical to
those found in the model without a modulus.  As $\eps_w$ approaches its
critical value \eref{epsbound} the results for the spectral index and
power spectrum change rapidly, and inflation breaks down abruptly.

\section {Conclusions}

In this paper we combined the hybrid inflation model of \cite{antusch}
with a KKLT like modulus sector.  The inflaton mass is protected by a
shift symmetry, and remains massless (at tree level) even in the
presence of the modulus sector.  This is in sharp contrast with
standard SUGRA hybrid inflation.  

The vacuum after inflation and the inflationary trajectory are
corrected by the modulus sector.  These corrections are under control
and do not disrupt inflation provided the modulus sector satisfies the
constraint $D_T W_{\rm mod} \approx W_{\rm mod} \approx 0$.  The first
condition is automatic in the minimum of the potential, the second
conditions can be satisfied by fine-tuning the parameters in the
potential. This is the same fine-tuning needed to get a hierarchy
between the gravitino and modulus mass, and which allows for low scale
SUSY breaking yet high scale inflation.  As explicit examples, the
original KKLT modulus stabilization scheme \cite{kklt} does not
satisfy the above condition, whereas the fine-tuned model of Kallosh
\& Linde \cite{kl1,kl2} does. 

Why inflation works for a modulus sector with small scale SUSY
breaking can be easily understood by considering the relevant scales
in the system.

\begin{enumerate}

\item The modulus mass $m_T \propto W_0$ which sets the height of the
barrier in the modulus potential.  It has to be larger than the
inflationary scale for the modulus to be stabilized during inflation;
this implies the condition  \eref{constraint1}.

\item The energy density during inflation $V_{\rm inf} = \bar \kappa^2
M^4$, which determines the size of the density perturbations. To get
the observed amplitude we find $V_{\rm inf}^{1/4} \sim M_{\rm GUT}$ is of
the order of the grand unified scale.  

\item The vacuum gravitino mass $m_{3/2} = \e^{K/2}|W| \propto
\eps_W$, which sets the scale of the moduli corrections to the
inflationary potential. It cannot be too large, the bound
\eref{epsbound} translates in a bound on the gravitino mass $m_{3/2}
\lesssim 10^9-10^{10}{\rm GeV}/\sigma_0^{3/2}$.
\end{enumerate}

In summary, we find that it is possible to extend the, in itself
already very promising, model of supersymmetric hybrid inflation
proposed in Ref. \cite{antusch} with a moduli sector. It is absolutely
necessary to have a modulus sector that does not break SUSY too
badly. Therefore we need to tune the parameters in the
superpotential. As a bonus, however, we find that our extended model
can accommodate TeV-scale SUSY breaking.

\section*{Acknowledgments}
The authors are supported by a VIDI grant from the Dutch Science Organization (NWO).

\appendix
\section{Coleman-Weinberg potential}
\label{A:CW}

The Coleman-Weinberg potential is \cite{CW}
\be
V_{\rm CW} = \frac1{64\pi^{2}} \sum_{i}(-1)^{F} m_{i}^{4}
 \ln\(\frac{m_{i}^{2}}{\Lambda^{2}}\)
\label{CW}
\ee
where the sum is over all masses, with $F=1$ for bosons and $F=-1$ for
fermions, and $\Lambda$ is the cutoff scale.  Only the
$n_{i}$-dependent mass terms lift the inflaton potential, in our case
these are the waterfall field masses and their fermionic partners. They
can be written in the form
\be
m^{2}_{h_{r,i}} = \mu (x^2 +y^2  \pm 1),
\qquad
\tilde m^{2}_{h_{r,i}} = \mu^2 x^2
\label{massh}
\ee
with $\mu,x,y$ given for inflation without and with a modulus field
respectively by (\ref{xy1},\ref{xy2}). The waterfall field $h_{r}$
becomes tachyonic and inflation ends for $x^{c} =1$.

Even though $y^{2} \ll x^{2}$ is clearly subdominant in the
expression for the mass terms \eref{massh}, they are important for the
shape of the loop potential.  This is because the dominant
contributions of the boson mass cancels with that of the fermion mass
in \eref{CW}. The loop potential becomes
\ba
V_{\rm CW}
=\(\frac{\mu^4}{32\pi^{2}}\) &\bigg[&
2(1+x^{2}y^{2} +y^{4}) \ln\( \frac{x^{2}\mu^{2}}{Q^{2}}\)  \\
&+&(x^{2}+y^{2}+1)^{2} \ln\(1+\frac{y^{2}+1}{x^{2}} \)
+(x^{2}+y^{2}-1)^{2} \ln\(1+\frac{y^{2}-1}{x^{2}} \)\bigg],\nn
\label{Vloop}
\ea
with $Q$ the renormalization scale which we fix to $Q = \mu = \tilde m_{h_{r,i}}|_{x=1}$. 
For negative values $y^2 <0 $ the potential develops a maximum at
large $x$.  Inflation has to take place on the left of the maximum,
for the inflaton field to roll towards the ``right'' minimum.  This
also means that if the maximum is to close to the critical value, it
is impossible to get 60 e-folds of inflation.  To see the maximum
appearing, we can take the large $x$ limit of the potential
\be
\lim_{x\to \infty} \(\frac{32\pi^{2}}{\mu^{4}}\) V_{\rm CW}
= 3 + 4 \ln(x) + 2y^{2}x^2\big(1+4 \ln(x)\big).
\ee
The slope of the potential at large $x$ gets a positive contribution
from the $y^{0}$-term, and a positive or negative contribution from
the $y^{2}$ correction; if the latter is negative, the potential has a
maximum.  The slope is
\be
\lim_{x\to \infty} \partial_x V_{\rm CW}
= \frac{4}{x} +  4xy^{2}(3+4 \log( x)),
\ee
which vanishes for
\be
x^{2}_{\rm max} = -(y^{2}(3+4\ln(x_{\rm max}))^{-1} 
\label{xmax}
\ee
which is only a solution for $y^2<0$.  Numerically we find for
$\kappa_{SH} = \O(1)$ that $x_{\rm max} =50 - 100$ in the absence of
moduli corrections (i.e. using \eref{xy1}).


\end{document}